\documentclass[sigconf, nonacm]{acmart}

\usepackage{pvldb}

\usepackage{xspace}
\usepackage{enumitem}
\usepackage{amsthm}

\usepackage[linesnumbered,ruled,vlined,boxed,noend]{algorithm2e}
\usepackage{booktabs}
\usepackage{multirow}
\usepackage{amsmath}
\usepackage{pifont}
\usepackage{colortbl}
\usepackage{xcolor}
\usepackage{array}
\usepackage{tabularx}

\usepackage{listings}
\theoremstyle{definition}
\newtheoremstyle{grbenchexample}
  {0.35em}                
  {0.35em}                
  {\itshape}              
  {0pt}                   
  {\bfseries\itshape}     
  {.}                     
  {0.5em}                 
  {}                      

\theoremstyle{grbenchexample}
\newtheorem{example}{Example}

\definecolor{queryblue}{RGB}{25,55,210}
\definecolor{querymagenta}{RGB}{190,25,205}
\definecolor{querygray}{RGB}{115,115,115}

\lstdefinelanguage{SQLPGQ}[]{SQL}{
  alsoletter={_},
  morekeywords={
    GRAPH_TABLE,
    MATCH,
    COLUMNS,
    ANY,
    SHORTEST,
    CHEAPEST,
    COST,
    PATH_LENGTH,
    ELEMENT_ID
  },
  sensitive=false
}

\lstdefinestyle{grbenchquery}{
  language=SQLPGQ,
  basicstyle=\ttfamily\fontsize{7.2}{8.4}\selectfont,
  keywordstyle=\color{queryblue},
  stringstyle=\color{querymagenta},
  commentstyle=\color{querygray}\itshape,
  identifierstyle=\color{black},
  numbers=left,
  numberstyle=\ttfamily\fontsize{5.8}{6.8}\selectfont
              \color{querygray},
  numbersep=5pt,
  stepnumber=1,
  xleftmargin=1.25em,
  frame=none,
  showstringspaces=false,
  columns=fullflexible,
  keepspaces=true,
  breaklines=true,
  breakatwhitespace=false,
  tabsize=2,
  aboveskip=0.25em,
  belowskip=0.45em
}

\newcommand{\myparagraph}[1]{\vspace{0.3\baselineskip}\noindent{\textbf{#1.}}~}

\newcommand{\bench}{\texttt{GRBench}\xspace}

\AtBeginDocument{%
  }

\renewcommand\vldbdoi{XX.XX/XXX.XX}
\renewcommand\vldbpages{XXX-XXX}
\renewcommand\vldbauthors{Zepeng Liu, Xinxin Huang, Xuanming Liu, Sheng Wang, Zhiyong Peng}
\renewcommand\vldbvolume{18}
\renewcommand\vldbissue{11}
\renewcommand\vldbyear{2027}
\renewcommand\vldbavailabilityurl{https://github.com/whu-totemdb/GRBench}

\begin{document}

\title{\bench: A Comprehensive Benchmark Evaluation for Graph-relational Data Management}
\author{Zepeng Liu$^1$, Xinxin Huang$^1$, Xuanming Liu$^1$, Sheng Wang$^{1}$, Zhiyong Peng$^{1,2}$}
\affiliation{%
  \institution{$^1$School of Computer Science, Wuhan University, $^2$Big Data Institute, Wuhan University}
  \streetaddress{}
  \city{}
  \country{}
}
\email{[liuzp\_063, xinxinhuang976, Teddy233, swangcs, peng]@whu.edu.cn}

\begin{abstract}
Modern data-intensive applications increasingly require database systems to manage structured records and graph data. This demand gives rise to \emph{graph-relational data management}, spanning storage, query processing, and optimization across relational and graph data. In response, relational database extensions, multi-model databases, and dedicated graph-relational systems have emerged with diverse architectures. However, evaluation methodologies have not kept pace. Existing relational and graph benchmarks assess the two models largely in isolation, while multi-model benchmarks provide limited coverage of graph-relational workloads. Available graph-relational workloads mainly support functional validation and end-to-end latency measurement, revealing little about how storage, operator, and optimization designs affect performance.
To evaluate system capabilities in graph-relational data management, we present \bench. First, \bench constructs a linked graph-relational schema from the real-world SciSciNet-v2 dataset and derives scalable instances through consistency-preserving subset extraction. Second, it organizes purpose-built query series for controlled evaluation of query processing and system components. Third, \bench provides semantically equivalent native query formulations and evaluates representative system architectures through a unified, multidimensional methodology. Based on this evaluation, we analyze design trade-offs and identify open challenges to guide future system design and optimization.
\end{abstract}

\maketitle

\vldbtopmatter

\section{INTRODUCTION}
Graph data~\cite{hogan2021knowledge,besta2024demystifying} has become widely used for representing entities and their relationships, while structured attributes and records remain naturally managed in relational form. Modern data-intensive applications \cite{guo2024multimodel,deutsch2022graph,gheerbrant2025gql,priem2022openalex} increasingly require database systems to support relational and graph data within the same workload. 

For example, a scholarly analyst may ask, \emph{``which papers published after 2020 by authors from a target institution are within one citation hop of a given paper?''} Answering this query requires relational filtering and joins over publication and author metadata with graph pattern matching over citation relationships, where relational results constrain graph search. We refer to such queries as \emph{graph-relational hybrid queries (GRHQs)}, which combine relational and graph operations in one query.
This demand gives rise to \emph{graph-relational data management}, requiring systems to support relational data management and graph processing and, more importantly, efficiently execute and optimize GRHQs.

\myparagraph{Motivations and Research Gaps}
To date, graph-relational data management has been supported by a growing range of database extensions \cite{duckpgq,apache}, multi-model databases \cite{ArangoDB2025,ritter2021orientdb}, and dedicated graph-relational systems \cite{jin2022grain,jin2022graindb,luo2025relgo,lee2024chimera,he2025graphiti}. The proliferation of such systems raises a natural research question: \emph{Is there a unified benchmark that can systematically evaluate the capabilities of these systems for graph-relational data management?}

Existing relational benchmarks, including Transaction Processing Performance Council (TPC) workloads \cite{tpc-c,tpc-h,tpc-ds} and optimizer benchmarks such as the Join Order Benchmark (JOB) \cite{leis2015job}, are valuable for studying transaction processing, analytics, join optimization, and cardinality estimation \cite{han2021cardinality}, but do not evaluate graph-specific query processing, such as graph pattern matching.
Conversely, graph benchmarks, including the Linked Data Benchmark Council (LDBC) suites \cite{erling2015ldbc,sz2022ldbc,iosup2016graphalytics,qi2025finbench}, primarily target graph-native transactional, analytical, and algorithmic workloads, without explicitly evaluating the interaction between relational joins and graph operations.
Multi-model benchmarks broaden the evaluation scope to heterogeneous data management \cite{ghazal2013bigbench,zhang2019unibench,kim2022m2bench}, but prioritize broad model coverage over the interaction between relational and graph processing, resulting in limited support for graph-relational workloads.
As a result, existing benchmarks largely evaluate relational and graph capabilities in isolation, or provide only limited support for the systematic evaluation of graph-relational data management.
We summarize the research gaps as follows: 
\begin{itemize}[left=0pt]
    \item \textbf{Gap 1}: Single-model benchmarks evaluate relational and graph processing in isolation. Even when relational and graph benchmarks are used together, they cannot capture the cross-model interactions and therefore cannot evaluate graph-relational data management as an integrated workload.
    \item \textbf{Gap 2}: Multi-model benchmarks provide almost no workloads dedicated to GRHQs. None of UniBench's 10 queries \cite{zhang2019unibench} and only one of M2Bench's 17 tasks \cite{kim2022m2bench} jointly exercise relational and graph processing. Such sparse coverage is insufficient to systematically characterize performance differences across systems.
    \item \textbf{Gap 3}: For the few available graph-relational workloads, existing benchmarks primarily support functional validation and end-to-end latency measurement, but lack controlled query series for fine-grained evaluation of graph-relational storage designs, operator implementations, and GRHQ optimizations.
\end{itemize}


\myparagraph{Our Contributions}
To address these gaps, we propose \bench, a comprehensive benchmark for graph-relational data management. 
\underline{To address \textbf{Gap 1}}, \bench constructs an integrated graph-relational schema from SciSciNet-v2~\cite{lin2023sciscinet}, a large real-world science-of-science dataset, and derives multiple scale factors through consistent subset extraction. Relational tuples and graph entities are linked through common identifiers and primary/foreign-key constraints, enabling queries to combine relational attribute access with graph traversal.
\underline{To address \textbf{Gap 2}}, \bench defines 24 GRHQs, each instantiated into three variants preserving the same logical structure while varying predicate selectivity, yielding 72 instances. This systematically parameterized workload expands GRHQ coverage and enables controlled characterization of performance differences across selectivity regimes.
\underline{To address \textbf{Gap 3}}, \bench organizes its workload into purpose-built query series, each designed to stress a distinct layer of graph-relational processing. Through controlled variations in query structure, operator composition, data scale, and equivalent graph and relational formulations, these query series help isolate the effects of system components rather than exposing only aggregate system-level performance. This design enables fine-grained evaluation of storage, operator, and optimization choices, extending the benchmark beyond functional validation and end-to-end latency.
As systems expose different graph-relational query languages and syntax, we further provide semantically equivalent formulations tailored to each system and evaluate query language conciseness.
This paper makes the following contributions:
\begin{itemize}[leftmargin=*]
    \item We present \bench, a comprehensive benchmark built on an integrated real-world graph-relational schema, consistently scaled datasets, and purpose-built query series.
    
    \item We systematically characterize representative systems on graph-relational data management in terms of their storage architectures, execution models, and optimization techniques.
    
    \item We conduct a cross-system evaluation using unified metrics and semantically equivalent queries tailored to different query languages and system interfaces.
    
    \item We identify key design trade-offs and open challenges in GRHQ optimization, graph-aware execution, update consistency, and cross-language usability.
\end{itemize}

\section{RELATED WORK}
To the best of our knowledge, no existing benchmark has been explicitly designed to evaluate graph-relational data management in a systematic manner. We therefore review representative relational, graph, and multi-model benchmarks, focusing on their workload coverage and limitations for graph-relational evaluation.

\myparagraph{Benchmarks for Relational and Graph Databases}
Relational benchmarks cover workloads including online transaction processing (OLTP), online analytical processing (OLAP), and optimizer evaluation. TPC-C \cite{tpc-c} evaluates OLTP through order-entry transactions. TPC-H and TPC-DS \cite{tpc-h,tpc-ds} evaluate decision support and retail analytics with scalable generators, and the Star Schema Benchmark \cite{patrick2009ssb} distills TPC-H into a star-schema workload for data warehouse engines. These standardized benchmarks provide reproducible workloads for comparing transaction processing, scans, joins, and aggregations. Their synthetic data follows predefined distributions and is less suited to optimizer robustness under real-world correlations and skew. Real-data optimizer benchmarks address this gap by focusing on real-world datasets and query patterns. For example, JOB \cite{leis2015job} uses the IMDB data \cite{imdb} and 113 multi-join SQL queries to reveal join ordering and cardinality estimation errors. Han et al. \cite{han2021cardinality} extend this work using the STATS dataset \cite{statsdataset}, providing a broader workload measuring estimation accuracy and end-to-end execution. These benchmarks move relational evaluation beyond standardized OLTP/OLAP benchmarking toward optimizer robustness under real-world conditions.

The LDBC series provides graph database benchmarks covering distinct aspects of graph processing. 
LDBC SNB Interactive \cite{erling2015ldbc} emphasizes transactional graph processing, including short reads, complex graph pattern matching, and updates. 
LDBC SNB Business Intelligence \cite{sz2022ldbc} targets graph OLAP over a social-network dataset, with aggregation-heavy queries involving graph pattern matching, traversal, and connectivity tests. However, these queries operate over a single logical graph dataset or its relational encoding, rather than coordinating relational and graph representations.
For algorithmic use cases, LDBC Graphalytics \cite{iosup2016graphalytics} evaluates kernels such as breadth-first search (BFS), PageRank \cite{malewicz2010pregel}, and connected components. LDBC FinBench \cite{qi2025finbench} targets financial graph transactions with temporal filters, path predicates, and read/write operations. These benchmarks, while comprehensive in their domains, primarily focus on graph-native processing and do not adequately cover relational and graph operations within one system \cite{sz2022ldbc,qi2025finbench,pacaci2022streaming}.

\myparagraph{Benchmarks for Multi-model Databases}
{Multi-model benchmarks extend beyond a single data model to evaluate systems that support multiple data models. For instance, BigBench \cite{ghazal2013bigbench} adapts the TPC-DS retail benchmark to include semi-structured (e.g., web click logs) and unstructured data (e.g., text reviews), focusing on large-scale analytics, including SQL analytics, machine learning, and text/pattern analysis. UniBench \cite{zhang2019unibench} simulates an e-commerce scenario, combining relational, document, and graph data to evaluate cross-model joins, aggregations, and transactional consistency. M2Bench \cite{kim2022m2bench}, targeting relational, document, graph, and array models across e-commerce and healthcare, involves tasks combining at least two data models. 
While these benchmarks advance multi-model system evaluation, none is specifically designed to evaluate graph-relational workloads, where relational operations and graph processing interact within the same query execution. Although few tasks jointly exercise relational and graph processing, their limited coverage and lack of controlled query variants are insufficient to characterize graph-relational performance.}

\myparagraph{Adjacent Benchmarks on Heterogeneous Data}
{Recent benchmarks also consider heterogeneous or relationally connected data. FDABench~\cite{wang2026fdabench} evaluates data agents over structured and unstructured sources, while RelBench~\cite{robinson2024relbench} represents relational tables connected through primary and foreign keys as graphs for predictive tasks. These benchmarks target agent-based data analysis or predictive learning rather than execution and optimization of relational and graph operations within database systems.}

\section{DATA MODEL AND EVALUATION SCOPE}
This section sets the modeling assumptions and evaluation boundary used by \bench. We treat relational tables and property graphs as distinct but connected representations, where relational data captures structured records and graph data captures relationship-centric topology. This separation lets \bench evaluate how systems support interaction across the two representations within the query processing core of graph-relational data management.

\subsection{Relational and Graph Representations}

We consider two separate data representations. The relational representation stores structured entities and attributes in tables. Formally, a system contains a set of tables $\mathcal{R}=\{R_1, R_2, \dots, R_n\}$. Each table $R_i$ has a schema $\textit{attr}(R_i)$ and stores tuples over these attributes. In \bench, relational query fragments are expressed through selection, projection, and join. Among them, join is the main mechanism for linking records across relational tables.

The graph representation stores entities and their relationships as a property graph. We denote it as $G=(V,E,\lambda,\phi)$, where $V$ is the set of vertices, $E \subseteq V \times V$ is the set of edges, $\lambda$ assigns labels to vertices and edges, and $\phi$ maps each vertex or edge to a set of properties. In this representation, entities are modeled as vertices and semantic associations are modeled as edges. Graph queries mainly rely on graph-specific operators, including pattern matching and shortest-path search \cite{deutsch2022graph,martens2023paths,anadiotis2023connection}.

Although these representations remain distinct, they can be connected through shared identifiers. In \bench, relational tuples and graph vertices may reference each other through foreign keys, so relational tuples can be joined with graph vertices, while graph query results can be joined back with relational records through the same identifiers. This key-based correspondence provides the basis for relational-graph interaction in the benchmark and reflects a common pattern where structured records and graph entities are maintained separately but linked through entity identifiers \cite{nadal2021federated,papastefanatos2022rdfschema}.

\subsection{Evaluation Dimensions}\label{sec:Evaluation Dimensions}


\myparagraph{GRHQ Capability}
As the core dimension of \bench, it measures whether a system can execute GRHQs combining relational selection, projection, and joins with graph pattern matching in one logical task. It further examines whether the system supports relational-graph interaction and whether its optimizer can exploit information from both models to optimize GRHQs.

\myparagraph{Operator Scalability}
Most non-graph-native systems \cite{tenwolde2023duckpgq,apache,duckdb} prioritize pattern matching when adding graph query support. We therefore examine whether such systems can provide effective compatibility and extension strategies for other common graph operators, such as shortest-path search, and how these operators perform when they participate in hybrid queries with relational joins. This dimension also tests whether the system can optimize such workloads when relational predicates determine path endpoints and graph expansion must be coordinated with relational filtering.

\myparagraph{Pure Graph Query Capability}
This dimension isolates graph operator efficiency by evaluating graph pattern matching without relational joins. By removing relational composition while retaining comparable graph structures and patterns, \bench examines graph-specific operators and topology access paths without conflating them with graph-relational coordination or optimization. These graph-only queries provide controlled baselines for GRHQ workloads, helping determine whether bottlenecks arise from graph operator execution or relational-graph interaction. Including these baselines ensures that pure graph and hybrid queries are evaluated under consistent data, schema, and execution settings \cite{kim2021versatile,gupta2021columnargraph}.

\myparagraph{Graph Update Support}
Graph updates include vertex and edge insertions and deletions. Across database systems that support relational and graph processing, graph records and topology are maintained using different storage organizations, such as edge tables~\cite{jindal2014vertexica,apache,tenwolde2023duckpgq,gupta2021columnargraph}, adjacency indexes~\cite{hassan2018extending,jin2022grain,pandey2021terrace,teseo2021dynamicgraph}, and explicit source and target vertex identifiers~\cite{apache,ArangoDB2025}. Update workloads therefore reveal the cost of maintaining consistency across data records, topology structures, and auxiliary indexes, rather than merely the raw cost of inserting or deleting an individual record \cite{yu2024lsmgraph,shi2024spruce}.

\myparagraph{Relational-equivalent Baseline}
We compare selected graph queries and GRHQs with semantically equivalent join-based relational translations. This comparison evaluates whether graph-oriented operators and execution frameworks provide measurable benefits over mature relational execution for graph and hybrid workloads, thereby identifying when relational operators and optimizations are sufficient and when graph processing is necessary.

\subsection{Scope Boundaries}

\bench focuses on the query processing core of graph-relational data management. It targets centralized database workloads where structured records and graph relationships are queried together, with basic graph updates included to characterize the cost of topology maintenance. The current scope excludes highly concurrent OLTP workloads, distributed execution, aggregation-heavy hybrid analytics, and graph algorithm workloads.

\section{\texttt{GRBENCH}}

\subsection{Benchmark Overview}

\begin{figure*}[t]
    \centering
    \includegraphics[width=0.65\textwidth]{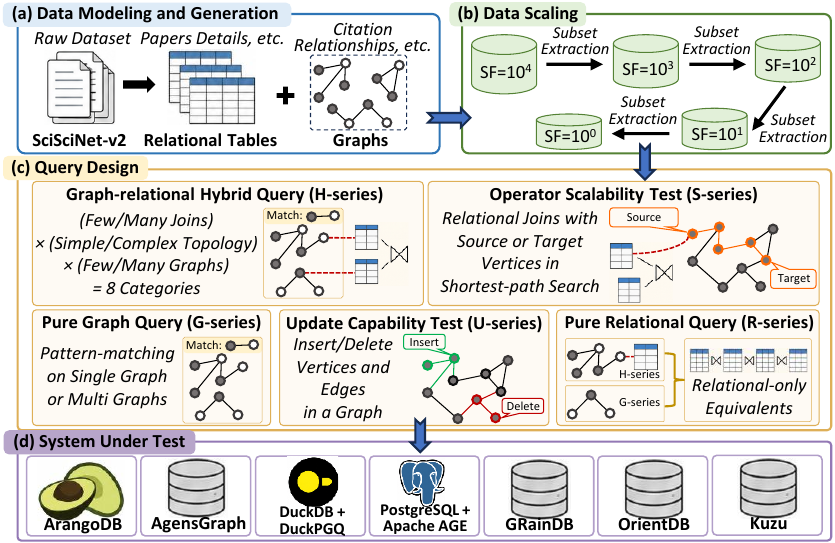}
    \vspace{-0.75em}
    \caption{An overview of \bench.}
    \vspace{-1.25em}
    \label{fig: overview}
\end{figure*}

\bench follows the pipeline shown in Figure~\ref{fig: overview}. We first construct a graph-relational benchmark instance from SciSciNet-v2 \cite{sciscinet,lin2023sciscinet}, as illustrated in Figure~\ref{fig: overview}(a). The raw scientific data is transformed into two coordinated representations: relational tables that store entity attributes and structured records, and graphs that capture citation, collaboration, and other relationship-centric structures. In this way, the benchmark focuses on the interaction between relational tables and graphs without introducing additional data models. Since the original dataset is already sufficiently large, \bench uses subset extraction rather than synthetic data generation for data scaling, as shown in Figure~\ref{fig: overview}(b). Different scale factors are obtained by extracting consistent subsets from the full dataset, while preserving the correspondence between the relational and graph parts.

Based on the scaled data instances, \bench organizes its workload into five query series, as summarized in Figure~\ref{fig: overview}(c). The H-series contains GRHQs, combining few/many joins, single/multiple matched subgraphs, and single-/multi-hop graph patterns to evaluate the core dual-model processing capability. The S-series targets operator scalability, especially relational joins over source or target vertices involved in shortest-path search. The G-series contains pure graph pattern matching queries over one or multiple matched subgraphs, serving as the graph-only baseline. The U-series evaluates update capability through vertex and edge insertions/deletions. The R-series provides relational-only equivalents of all queries in H- and G-series, helping isolate the relational execution cost. 
Finally, \bench evaluates representative systems as shown in Figure~\ref{fig: overview}(d).


\subsection{Dataset and Schema Design}

\myparagraph{Dataset Selection}
\bench uses SciSciNet-v2 \cite{sciscinet,lin2023sciscinet}, a large-scale linked dataset for science-of-science research, as its data source. The dataset describes the scholarly publication ecosystem and contains major entities such as papers, authors, and institutions, together with relationship records including authorship and citation links. The public release contains approximately 250M\footnote{We use K for Thousand, M for Million, and B for Billion in this paper.} papers, 100M authors, and 2.49B paper-reference records.

SciSciNet-v2 is well aligned with graph-relational benchmarking for the following reasons:
\begin{itemize}[leftmargin=*]
    \item \textit{Large scale.} The dataset is large enough to support non-trivial subset extraction and to stress storage, indexing, and query execution over both relations and graphs.
    \item \textit{Complex real-world distributions.} It preserves skewed degree distributions, correlated attributes, and long-tail scientific relationships, making it more representative than synthetic generators for optimizer and execution studies.
    \item \textit{Rich join and graph settings.} Structured metadata can be joined by paper, author, and affiliation identifiers, while citation, authorship, and external-link records form graph topologies. Together, they enable workloads where relational predicates and joins interact with graph pattern matching or traversal.
    \item \textit{Open availability.} The dataset is publicly available, improving reproducibility and avoiding licensed-data dependencies.
\end{itemize}

\begin{figure}[t]
    \centering
    \includegraphics[width=\columnwidth]{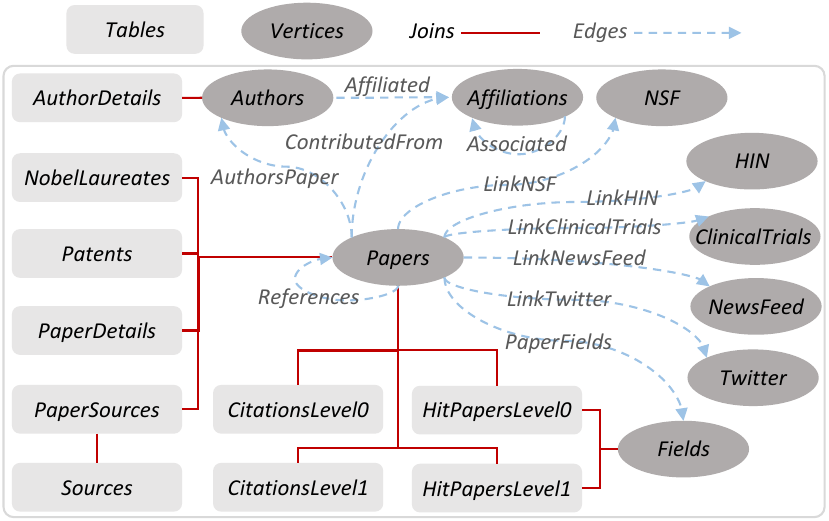}
    \vspace{-2em}
    \caption{Schema design of \bench.}
    \vspace{-1.75em}
    \label{fig: schema}
\end{figure}

\myparagraph{Schema Design}
We organize SciSciNet-v2 into the coupled schema shown in Figure~\ref{fig: schema}. The design keeps descriptive attributes in relational tables and represents relationship-centric records as property graphs. Concretely, the schema contains 10 relational tables, 9 vertex labels, and 11 edge-labeled property graphs, where each distinct edge label in the figure corresponds to one graph. Papers and authors act as the main anchors between the two models. Relational tables provide paper, author, and hit-paper attributes, while graph edges capture authorship, citation, and field membership.

The red solid lines in Figure~\ref{fig: schema} denote the relational join backbone. In total, the schema supports up to 12 schema-level joins, connecting metadata tables to their corresponding paper and field entities. This organization exposes the intended graph-relational interaction without flattening the dataset into a single representation. Relational predicates can restrict structured records, and graph patterns can then be evaluated over the aligned property graphs through shared identifiers.

\subsection{Data Scaling}

\bench derives five scale factors, ranging from $SF=10^0$ to $SF=10^4$, from SciSciNet-v2 through subset extraction rather than synthetic data generation, as shown in Figure~\ref{fig: overview}(b). The resulting instances span from 488K to 249M papers, from 74.1K to 100M authors, and from 249K to 2.49B reference edges, providing graph-relational workloads over a broad range of data scales. All retained records originate from the original dataset, so attribute values, key correlations, degree skew, and long-tail relationship patterns remain grounded in real data.
For each entity type, we define a per-step extraction ratio \(\rho\). For large tables and graphs, we set \(\rho=0.1\), yielding an approximately one-order-of-magnitude reduction between adjacent scale factors. Smaller external-evidence entities use larger ratios to avoid undersized instances. For example, \textit{Associated} and \textit{NobelLaureates} use \(\rho=0.4\), whereas the fixed field taxonomy is retained in full with \(\rho=1.0\). Consequently, even at \(SF=10^0\), all scalable entities contain thousands of records, while the field taxonomy remains fixed at 303 entries.


The extraction is performed jointly over the relational and graph components. We first sample anchor structures, such as paper references, and then derive the required paper, author, field, affiliation, and external-evidence entities from the retained identifiers. Dependent edges and relational tables are filtered using the same identifiers or deterministic key-based sampling over join attributes. This procedure preserves referential integrity and joinability across the two representations while making the scale factors reproducible. It also maintains graph consistency by retaining only edges with valid source and target identifiers and constructing graph projections from incident vertices, thereby avoiding isolated vertices in the evaluated graphs. Meanwhile, metadata required for relational joins remains available as structured tables or vertex attributes.

\subsection{Workload Design}

\bench organizes its workload into five query series, as summarized in Table~\ref{tab:workload-design}. All series share the same schema and scaled instances, allowing performance differences to be interpreted through controlled changes in query structure. The examples below use compact SQL/PGQ skeletons from the executable workload. They retain the core filtering, join, pattern matching, shortest-path search, and update operations. For each H-, S-, and G-series query template, we define three predicate-selectivity variants, denoted by the suffixes \texttt{a}, \texttt{b}, and \texttt{c}. Averaged across all query templates, the selectivities of the three variants are 45.19\%, 33.15\%, and 4.01\%, respectively. The corresponding R-series translations retain the same suffixes.

\myparagraph{H-series: Graph-relational Hybrid Queries}
The H-series follows a $2\times2\times2$ design over relational-join intensity (few or many), the number of matched subgraphs (single or multiple), and graph pattern depth (single- or multi-hop). This factorial design enables controlled analysis of how relational joins interact with graph pattern matching. For instance, Example~\ref{example:h} instantiates the few-join, multiple-subgraph, single-hop design cell with two matched subgraphs and two joins: one join combines the two graph query results, and the other associates the combined result with a relational table. The optimizer may vary when to apply the vertex predicates, the order in which to evaluate the two graph patterns, and when to perform the graph query result and relational joins. These choices can substantially affect intermediate result cardinalities, making the query a compact test of whether the optimizer can apply selective predicates early and coordinate graph-derived results with relational execution.

\begin{example}\label{example:h}
Retrieve papers with fewer than 150 citations that are referenced by a selected paper, together with newsfeed and Nobel-laureate records associated with that paper.
\end{example}

\begin{table}[t]
    \centering
    \small
    \setlength{\tabcolsep}{2.2pt}
    \renewcommand{\arraystretch}{0.95}
    \caption{Organization of the \bench workload.}
    \vspace{-1em}
    \label{tab:workload-design}
    \begin{tabularx}{\columnwidth}{@{}>{\centering\arraybackslash}c >{\centering\arraybackslash}c >{\centering\arraybackslash}X@{}}
    \toprule
    \textbf{Series} & \textbf{Queries} & \textbf{Description} \\
    \midrule
    \multirow{8}{*}{H}
        & H1--H3   & Many joins / Multiple subgraphs / Multi-hop \\
        & H4--H6   & Many joins / Multiple subgraphs / Single-hop \\
        & H7--H9   & Many joins / Single subgraph / Multi-hop \\
        & H10--H12 & Many joins / Single subgraph / Single-hop \\
        & H13--H15 & Few joins / Multiple subgraphs / Multi-hop \\
        & H16--H18 & Few joins / Multiple subgraphs / Single-hop \\
        & H19--H21 & Few joins / Single subgraph / Multi-hop \\
        & H22--H24 & Few joins / Single subgraph / Single-hop \\
    \midrule
    S & S1--S5 & Shortest-path search combined with relational joins \\
    \midrule
    \multirow{4}{*}{G}
        & G1--G2 & Multiple subgraphs / Multi-hop \\
        & G3--G4 & Multiple subgraphs / Single-hop \\
        & G5--G6 & Single subgraph / Multi-hop \\
        & G7--G8 & Single subgraph / Single-hop \\
    \midrule
    \multirow{3}{*}{U}
        & U1--U2 & Vertex-only insertion/deletion \\
        & U3--U4 & Edge insertion/deletion \\
        & U5--U6 & Vertex and incident-edge insertion/deletion \\
    \midrule
    \multirow{2}{*}{R}
        & RH1--RH24 & Relational equivalents of H-series queries \\
        & RG1--RG8   & Relational equivalents of G-series queries \\
    \bottomrule
    \end{tabularx}
    \vspace{-1.5em}
\end{table}

\begin{lstlisting}[style=grbenchquery]
SELECT MIN(Citation.cited_id), MIN(Feed.newsfeed_id),
       MIN(NL.laureate_id)
FROM GRAPH_TABLE(
  sciscinet_graph
  MATCH (src IS Papers)-[r IS References]->
        (dst IS Papers)                    -- pattern matching
  WHERE src.paperid = 'W2122344208'
    AND dst.citation_count < 150           -- filtering
  COLUMNS (src.paperid AS src_id, dst.paperid AS cited_id)
) AS Citation
JOIN GRAPH_TABLE(
  sciscinet_graph
  MATCH (p IS Papers)-[n IS LinkNewsFeed]->
        (f IS NewsFeed)                    -- pattern matching
  COLUMNS (p.paperid AS paperid, f.newsfeed_id AS newsfeed_id)
) AS Feed
  ON Citation.src_id = Feed.paperid        -- graph-result join
JOIN NobelLaureates AS NL
  ON Citation.src_id = NL.paperid;         -- relational join
\end{lstlisting}

\myparagraph{S-series: Operator Scalability Queries}
The S-series evaluates whether database systems can extend their graph processing support beyond pattern matching to additional graph operators and execute these operators efficiently. We use shortest-path search as the representative operator because it is one of the most widely used graph operations beyond pattern matching and requires systems to support topology expansion beyond fixed graph patterns. The five templates cover bounded and unbounded shortest paths, the composition of path results with other query relations, and cost-based path search. Their variants modify endpoint constraints and path cases, enabling controlled analysis of operator efficiency under different search spaces and binding conditions. Example~\ref{example:s} combines a bounded \texttt{ANY} \texttt{SHORTEST} search over 1 to 4 hops with a single-hop contribution pattern and one join between the two graph query results. The source and destination predicates constrain the path endpoints, while the join composes the shortest-path result with paper--institution bindings produced by graph pattern matching. This combination evaluates whether a system can extend its graph-processing architecture beyond pattern matching to support shortest-path search and whether the operator can be executed efficiently when integrated with graph-derived relations.

\begin{example}\label{example:s}
Retrieve institutions with productivity above 8,000 that are reachable from institution \texttt{I4210136388} by a shortest association path of 1 to 4 hops and that have contributed papers.
\end{example}

\begin{lstlisting}[style=grbenchquery]
SELECT MIN(Contribution.institutionid)
FROM GRAPH_TABLE(
  sciscinet_graph
  MATCH path = ANY SHORTEST
        (src IS Affiliations)-[af IS Associated]->{1,4}
        (dst IS Affiliations)            -- shortest-path search
  WHERE src.institutionid = 'I4210136388'
    AND dst.productivity > 8000          -- endpoint filtering
  COLUMNS (dst.institutionid AS institutionid)
) AS Reachable
LEFT JOIN GRAPH_TABLE(
  sciscinet_graph
  MATCH (paper IS Papers)-[cf IS ContributedFrom]->
        (inst IS Affiliations)           -- pattern matching
  COLUMNS (
    paper.paperid AS paperid,
    inst.institutionid AS institutionid
  )
) AS Contribution
  ON Reachable.institutionid =
     Contribution.institutionid;         -- graph-result join
\end{lstlisting}

\myparagraph{G-series: Pure Graph Queries}
The G-series follows a $2\times2$ design over the number of matched subgraphs (single or multiple) and pattern depth (single- or multi-hop), with two query templates in each design cell. By excluding relational joins, this series evaluates graph query execution at the operator level and isolates the efficiency of graph pattern matching from graph-relational composition. Example~\ref{example:g} instantiates the single-subgraph, multi-hop design cell with one four-hop association pattern. The query requires the system to apply vertex filtering and execute a fixed-depth graph pattern entirely within graph processing. Its performance therefore reflects graph query operator efficiency and provides an operator-level baseline for the corresponding H-series queries.

\begin{example}\label{example:g}
Retrieve institutions that are exactly four association hops away from institutions whose identifiers begin with \texttt{I42101230}.
\end{example}

\begin{lstlisting}[style=grbenchquery]
SELECT MIN(G.dst_id)
FROM GRAPH_TABLE(
  sciscinet_graph
  MATCH (src IS Affiliations)-[e1 IS Associated]->
        (m1 IS Affiliations)-[e2 IS Associated]->
        (m2 IS Affiliations)-[e3 IS Associated]->
        (m3 IS Affiliations)-[e4 IS Associated]->
        (dst IS Affiliations)              -- pattern matching
  WHERE src.institutionid LIKE 'I42101230%'-- vertex filtering
  COLUMNS (dst.institutionid AS dst_id)
) AS G;
\end{lstlisting}

\myparagraph{U-series: Update Capability Queries}
The U-series contains six update tasks, each organized as a deletion--insertion pair over the same records. The tasks cover isolated vertex updates, edge updates, and dependency-aware vertex updates that also modify incident edges, enabling controlled evaluation of both direct modification cost and consistency maintenance overhead. Example~\ref{example:u} removes and restores the batch of paper--author edges listed in \texttt{grbench\_update5\_papers\_to\_authors}. Using the same batch keeps the update scope fixed, allowing deletion and insertion costs to be compared under identical data conditions. Graph pattern queries executed before and after each operation verify that the logical graph view reflects the update. The task evaluates whether a system can maintain edge records, endpoint references, and topology-related structures consistently and efficiently.

\begin{example}\label{example:u}
Remove a predefined batch of paper--author relationships and subsequently restore the same relationships.
\end{example}

\begin{lstlisting}[style=grbenchquery]
DELETE FROM Papers_to_Authors AS E
WHERE EXISTS (
  SELECT 1
  FROM grbench_update5_papers_to_authors AS U
  WHERE U.paperid = E.paperid AND U.authorid = E.authorid
);                                           -- edge deletion
INSERT INTO Papers_to_Authors
SELECT *
FROM grbench_update5_papers_to_authors;      -- edge insertion
\end{lstlisting}

\myparagraph{R-series: Relational-only Equivalents}
The R-series translates selected H- and G-series queries into relational-only formulations over vertex and edge tables. Each translation preserves the semantics of the original query. This paired design provides a controlled baseline for assessing whether each system's graph query processing design delivers measurable performance benefits over semantically equivalent join-based relational execution. 

For instance, Example~\ref{example:r-h} translates the GRHQ in Example~\ref{example:h} by replacing its two graph patterns with relational joins. The one-hop citation pattern becomes joins among \texttt{Papers} and \texttt{References}, while the paper--newsfeed pattern becomes joins through \texttt{LinkNewsFeed}. 

\begin{example}\label{example:r-h}
Retrieve the same papers, newsfeed records, and Nobel-laureate records as Example~\ref{example:h} using only relational joins.
\end{example}

\begin{lstlisting}[style=grbenchquery]
SELECT MIN(dst.paperid), MIN(NF.newsfeed_id),
       MIN(NL.laureate_id)
FROM Papers AS src
JOIN References AS R ON src.paperid = R.citing_paperid
JOIN Papers AS dst ON R.cited_paperid = dst.paperid
 AND dst.citation_count < 150
JOIN LinkNewsFeed AS LNF ON src.paperid = LNF.paperid
JOIN NewsFeed AS NF ON LNF.newsfeed_id = NF.newsfeed_id
JOIN NobelLaureates AS NL ON src.paperid = NL.paperid
WHERE src.paperid = 'W2122344208';
\end{lstlisting}

Example~\ref{example:r-g} translates the pure graph query in Example~\ref{example:g}. Its four-hop association pattern becomes an alternating sequence of joins over \texttt{Affiliations} vertices and \texttt{Associated} edges. Because the original query contains no relational composition, this pair directly compares graph pattern matching with join-based relational evaluation at the graph-query operator level.

\begin{example}\label{example:r-g}
Retrieve the same four-hop associated institutions as Example~\ref{example:g} using relational joins.
\end{example}

\begin{lstlisting}[style=grbenchquery]
SELECT MIN(a4.institutionid)
FROM Affiliations AS a0
JOIN Associated AS e1
  ON a0.institutionid = e1.institution_id
JOIN Affiliations AS a1
  ON e1.associated_institution_id = a1.institutionid
JOIN Associated AS e2
  ON a1.institutionid = e2.institution_id
JOIN Affiliations AS a2
  ON e2.associated_institution_id = a2.institutionid
JOIN Associated AS e3
  ON a2.institutionid = e3.institution_id
JOIN Affiliations AS a3
  ON e3.associated_institution_id = a3.institutionid
JOIN Associated AS e4
  ON a3.institutionid = e4.institution_id
JOIN Affiliations AS a4
  ON e4.associated_institution_id = a4.institutionid
WHERE a0.institutionid LIKE 'I42101230%';
\end{lstlisting}

\section{SYSTEM LANDSCAPE}\label{sec:survey}

This section reviews representative systems that support graph-relational data management, mainly focusing on where graph storage and execution are integrated in the original system.


\subsection{Relational Systems with Graph Extensions}

These systems retain relational storage and add graph interfaces or execution components above an existing relational engine.

\myparagraph{Vertexica \cite{jindal2014vertexica}}
Vertexica layers Pregel's vertex-centric computation model~\cite{malewicz2010pregel} over Vertica~\cite{lamb2012vertica}. It stores vertices, edges, and inter-superstep messages in three relational tables, while a stored procedure coordinator invokes parallel worker user-defined functions (UDFs). Graph algorithms are thereby compiled into SQL/UDF workflows and executed by the unmodified relational engine.

\myparagraph{PostgreSQL with Apache AGE \cite{PostgreSQL,apache}}
AGE creates a PostgreSQL namespace for each graph and stores vertex and edge labels in label-specific tables derived from parent vertex and edge tables. Cypher statements embedded in \texttt{cypher()} are parsed and transformed into PostgreSQL query tree nodes, while graph entities and properties are represented with \texttt{agtype}. These query trees are planned and executed by PostgreSQL. The returned \texttt{SETOF record} can be composed with ordinary SQL relations.

\myparagraph{DuckDB with DuckPGQ \cite{duckdb,duckpgq,tenwolde2023duckpgq}}
DuckPGQ defines property graphs over existing typed vertex and edge tables without duplicating the base data. The extension adds SQL/PGQ parsing, vectorized UDFs, and graph-specific data structures and algorithms that are injected into DuckDB plans. Fixed patterns reuse DuckDB's relational execution, while path queries can build an on-the-fly compressed sparse row (CSR) representation \cite{gupta2021columnargraph}.

\subsection{Graph-native Multi-model Systems}

These systems natively support the graph model alongside other models, sharing storage or execution infrastructure rather than exposing graph functionality as an extension to a relational engine.

\myparagraph{ArangoDB \cite{ArangoDB2025}}
ArangoDB stores vertices as documents in vertex collections and edges as documents in edge collections. Each edge carries \texttt{\_from} and \texttt{\_to} endpoint references, for which edge indexes are created automatically. ArangoDB query language (AQL) provides native traversal and path operators over named or anonymous graphs, with controls for direction, depth, and traversal strategy.

\myparagraph{AgensGraph \cite{agensgraph}}
Built on PostgreSQL's storage and transaction infrastructure, AgensGraph is a multi-model database supporting relational, document, and graph models. Property graphs use vertex and edge labels, with hierarchical label organization and indexes on both vertices and edges. SQL and openCypher can be used independently or combined within a single query, while both query modes share the same underlying database infrastructure. Both modes execute within the same PostgreSQL-derived engine.

\myparagraph{OrientDB \cite{ritter2021orientdb}}
OrientDB supports graph, document, key-value, and object models over a common record store. Regular edges are separate records linked from both endpoint vertices, whereas lightweight edges are stored as direct links inside vertex records. Its SQL dialect provides \texttt{MATCH} and \texttt{TRAVERSE} for link navigation, and indexed predicates can locate starting records before traversal.

\subsection{Native Graph-relational Systems}
Systems in this group integrate graph-specific storage and execution mechanisms directly into the database core, including graph-relational designs and K\`{u}zu as a native graph database reference.

\myparagraph{K\`{u}zu \cite{kuzu2023}}
K\`{u}zu is a disk-based, columnar graph DBMS based on a structured property-graph model and the Cypher query language. Node properties are stored in column files, while graph topology and relationship properties are double-indexed in forward and backward CSR adjacency structures. Its factorized query processor provides binary and worst-case-optimal join operators for many-to-many, cyclic, and recursive graph patterns.

\myparagraph{GRFusion \cite{hassan2018extending}}
Built into the in-memory relational engine of VoltDB \cite{stonebraker2013voltdb}, GRFusion introduces first-class graph views. Vertex and edge properties remain in relational storage, whereas graph topology is maintained in adjacency-list indexes. During query execution, \texttt{VertexScan}, \texttt{EdgeScan}, and \texttt{PathScan} produce tuples that are connected to relational operators through binding joins.

\myparagraph{GRainDB \cite{jin2022graindb,jin2022grain}}
Rather than materializing a separate graph store, GRainDB retains DuckDB's relational tables and columnar executor. For each declared relationship, system-level row identifiers are stored in extended tables, with optional adjacency-list-like RID indexes. Graph traversal is then evaluated through predefined RID-based joins over this metadata.

\myparagraph{RelGo \cite{luo2025relgo}}
Using relational tables as its physical storage layer, RelGo stores vertices and edges while exposing topology through RID-based graph indexes. Its execution model represents graph-relational queries as selection, projection, join, and match plans, retaining an explicit matching operator rather than immediately lowering graph patterns into joins. The resulting physical plan is encapsulated as a graph-table scan in DuckDB.

\myparagraph{Chimera \cite{lee2024chimera}}
In contrast to the relational-storage designs above, Chimera adopts a dual-store architecture that places topology in a native graph store and properties in PostgreSQL. A bijective mapping links graph records to relational tuples. The Traversal-Join operator combines topology traversal, relational joining, and graph-relational mapping within a single execution abstraction.

\subsection{Graph-relational Query Acceleration}


\myparagraph{Reuse of Relational Execution}
Vertexica obtains parallelism from Vertica's worker UDF execution, while AGE and AgensGraph mainly reuse the scans, joins, indexes, and planning infrastructure of their underlying relational engines. DuckPGQ further uses vectorized graph functions: bounded SQL/PGQ patterns are translated into relational plans, while path queries use vectorized path algorithms over an on-demand CSR representation. These approaches benefit from mature relational execution, although long or cyclic patterns may still expand into costly join plans \cite{tenwolde2023duckpgq,figueira2022datapath}.

\myparagraph{Topology-aware Graph Processing}
ArangoDB accelerates traversal through endpoint and vertex-centric indexes, traversal pruning, and configurable search strategies. OrientDB's \texttt{MATCH} executor selects indexed starting records before following graph links. GRFusion provides native scan and path operators, while GRainDB replaces value-based joins with predefined RID joins and propagates bitmap filters through sideways information passing (SIP). K\`{u}zu targets many-to-many and cyclic patterns through factorized execution with binary and worst-case-optimal joins.

\myparagraph{Cross-model Query Optimization}
RelGo retains graph matching in selection, projection, join, and match plans, applies graph-specific decomposition and rewrite rules, and maps matching operations to expansion and expansion-intersection operators before relational optimization. Chimera constructs Traversal-Join graphs and uses dynamic programming to enumerate plans that interleave graph traversal, relational joins, and graph-relational mappings. These designs optimize cross-model orderings instead of relying on graph-agnostic relational planning.

\section{EXPERIMENTS}

In this section, we systematically evaluate the dimensions introduced in Section~\ref{sec:Evaluation Dimensions} to answer the following research questions: 
\begin{itemize}[left=0pt]
    \item (\textbf{RQ1}) How do existing systems perform on GRHQs overall, and which design choices explain their efficiency advantages? 
    \item (\textbf{RQ2}) How extensible is graph query support, and can their architectures incorporate operations beyond pattern matching?
    \item (\textbf{RQ3}) How efficiently do existing systems execute graph pattern matching at the operator level? 
    \item (\textbf{RQ4}) How efficiently do systems process graph updates, and can they preserve consistency during updates? 
    \item (\textbf{RQ5}) Compared with equivalent relational execution, do graph query designs deliver measurable performance gains?
\end{itemize}
Furthermore, we examine how data scale and predicate selectivity affect query execution in Sections~\ref{sec:exp data scale}--\ref{sec:exp selectivity}, and evaluate query-language conciseness and resource utilization in Sections~\ref{sec:exp language}--\ref{sec:exp resource}.

\subsection{Experimental Setup}

\myparagraph{Systems Under Test}
Since not all systems surveyed in Section~\ref{sec:survey} are open source, we evaluate all open-source systems among them, as shown in Figure~\ref{fig: overview}(d). Specifically, we use PostgreSQL 15.6 with Apache AGE 1.6.0, DuckDB 1.2.2 with DuckPGQ (commit \texttt{97a0c65}), ArangoDB 3.12.4, AgensGraph 2.16, OrientDB Community 3.2.37, GRainDB (commit \texttt{be901b3}), and K\`{u}zu 0.11.3.


\begin{figure*}[t]
    \centering
    \includegraphics[width=\textwidth]{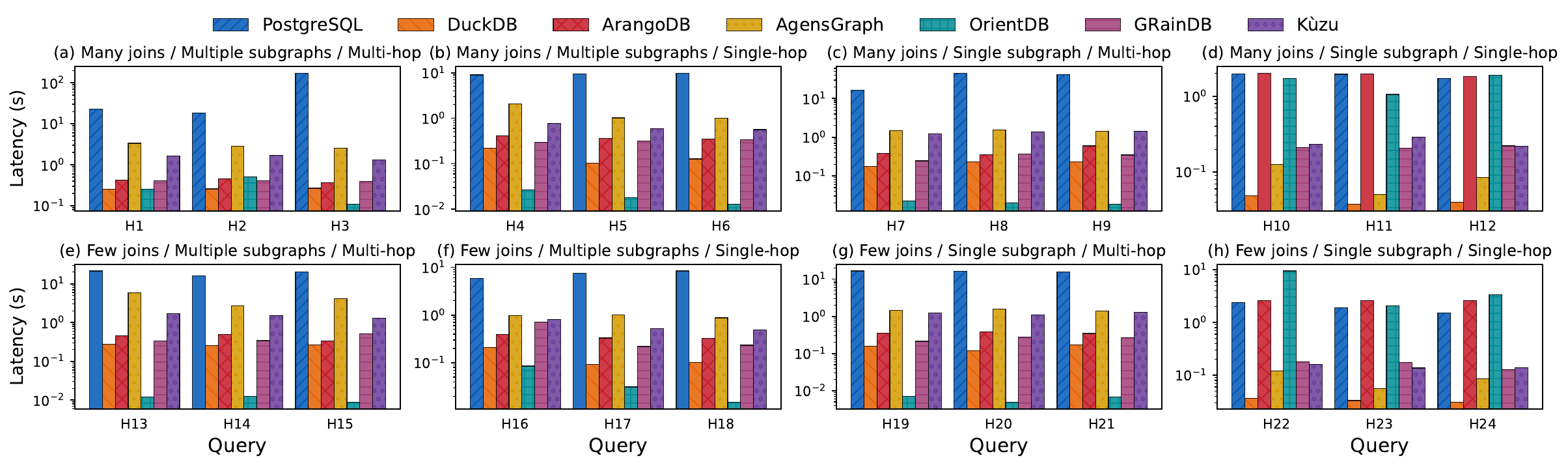}
    \vspace{-2.75em}
    \caption{Geometric mean latency across the three predicate-selectivity variants of each H-series query at $SF=10^1$.}
    \vspace{-1.5em}
    \label{fig:exp GRHQ}
\end{figure*}

\begin{figure*}[t]
    \centering
    \includegraphics[width=\textwidth]{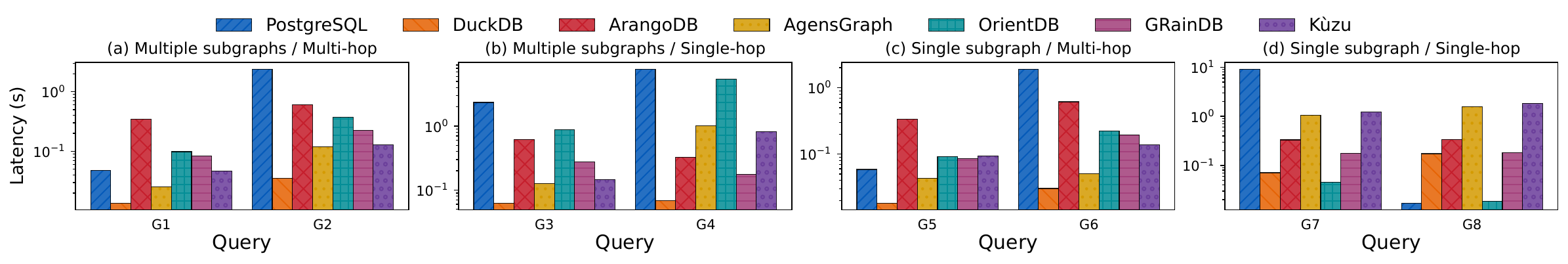}
    \vspace{-2.75em}
    \caption{Geometric mean latency across the three predicate-selectivity variants of each G-series query at $SF=10^1$.}
    \vspace{-1em}
    \label{fig:exp graph}
\end{figure*}

\myparagraph{Evaluation Metrics}
We evaluate all comparisons from:
\begin{itemize}[left=0pt]
    \item \underline{\textit{End-to-end Latency}}. 
    For a query $Q$, $\mathcal{L}(Q)$ denotes the wall-clock time from submission to complete result materialization. For a query set $C$, we report the geometric mean:
    \begin{equation}
        \mathcal{L}_{GEO}(C)=
        \exp\left(\frac{1}{|C|}\sum_{Q\in C}\ln \mathcal{L}(Q)\right).
    \end{equation}
    
    \item \underline{\textit{Graph-relational Speedup}}. 
    For an H- or G-series query $Q$ and its R-series equivalent $RQ$, we use $\mathcal{L}(RQ)/\mathcal{L}(Q)$ as the speedup. A value above one indicates faster graph-based execution.

    \item \underline{\textit{Query Language Conciseness}}. 
    We express each query in the native form supported by each system and quantify conciseness using Halstead-style complexity metrics~\cite{halstead1977elements}. After removing comments and normalizing formatting, a unified tokenizer counts distinct operators $n_1$, distinct operands $n_2$, and their total occurrences $N_1$ and $N_2$. We compute:
    \begin{equation}
    \begin{gathered}
        n=n_1+n_2,\quad
        N=N_1+N_2,\quad
        \mathcal{V}=N\log_2 n, \\
        \mathcal{D}=\frac{n_1}{2}\times\frac{N_2}{n_2},\quad
        \mathcal{E}=\mathcal{D}\times\mathcal{V},
    \end{gathered}
    \end{equation}
    where $\mathcal{V}$, $\mathcal{D}$, and $\mathcal{E}$ denote query volume, difficulty, and estimated effort, respectively. Lower values indicate more concise formulations for the same logical task.

    \item \underline{\textit{CPU and Memory Utilization}}. 
    For each query $Q$, we sample CPU and memory usage during execution and record their average and peak values. CPU is reported as equivalent utilized cores, where 100\% corresponds to one core, and memory in GB. These metrics distinguish efficient execution from low latency achieved through greater resource consumption.
\end{itemize}

\myparagraph{Implementation Details}
Experiments run on a server with two Intel Xeon Platinum 8352V CPUs (2.10\,GHz, 72 physical cores and 144 threads), 512\,GiB DDR4 memory, and Ubuntu 22.04.5 LTS.
All systems use default configurations in Docker containers without CPU or memory limits. DuckDB therefore uses 144 threads by default. Before execution, we construct system-specific indexes and auxiliary structures, including primary-key indexes on relational tables, adjacency structures, and predefined-join indexes for graph processing. For each system, we run the complete workload three times in fixed order without restarting or clearing database and operating system caches, and report median latency. Later repetitions benefit from a warmed execution environment. Each query is limited to five hours (18,000 seconds), with timeouts recorded as 18,000 seconds. For each query, we verify that all systems produce \textit{identical results} to ensure semantic equivalence across implementations.

\subsection{Efficiency on GRHQs (RQ1)}\label{sec:exp GRHQ}

We evaluate all seven systems on 24 H-series GRHQs and report their latencies in Figure~\ref{fig:exp GRHQ}. We make the following observations:
(1) PostgreSQL with AGE shows the highest latency in most H-series queries, especially with multiple graph subqueries or multi-hop patterns. AGE exposes Cypher execution through the set-returning \texttt{cypher()} interface, and graph results are composed through PostgreSQL joins. This separation between graph execution and relational composition introduces overhead when multiple graph-derived relations interact.
(2) OrientDB achieves the lowest latency in most queries, while DuckDB performs best in the remaining cases. OrientDB's advantage mainly appears with exact identifier predicates, where the starting vertex can be located before traversal through its \texttt{MATCH} execution and indexes.
(3) H10--H12 and H22--H24 show a different pattern. DuckDB achieves the lowest latency, while OrientDB and ArangoDB become less competitive. These queries use prefix predicates rather than exact identifiers, preventing direct binding of a starting vertex before traversal. DuckPGQ translates SQL/PGQ graph patterns into DuckDB plans for execution within the relational framework.
(4) DuckDB maintains consistently low latency across H-series structures. DuckPGQ lowers graph patterns into DuckDB's relational execution framework, allowing graph matching to reuse existing relational operators and benefit from its vectorized execution model.
(5) Join count, matched-subgraph count, and hop depth affect systems differently rather than producing a monotonic trend. PostgreSQL becomes increasingly costly with more graph-relational composition, while GRainDB remains relatively stable across query structures. Its predefined joins and RID-based access paths reduce repeated relationship expansion, showing that GRHQ efficiency depends on how systems integrate graph and relational execution.

\begin{table}[t]
\centering
\small
\setlength{\tabcolsep}{4.5pt}
\renewcommand{\arraystretch}{0.92}
\caption{S-series latency (seconds) at $SF=10^1$.}
\vspace{-1.25em}
\label{tab:scalability-s-sf10-1}
\definecolor{scalabilitybestgreen}{RGB}{180,205,180}
\definecolor{scalabilitysecondgreen}{RGB}{220,232,220}
\newcommand{\scalabilitybest}[1]{\cellcolor{scalabilitybestgreen}#1}
\newcommand{\scalabilitysecond}[1]{\cellcolor{scalabilitysecondgreen}#1}
\begin{tabular}{@{}cccccc@{}}
\toprule
\textbf{Systems} & \textbf{S1} & \textbf{S2} & \textbf{S3} & \textbf{S4} & \textbf{S5} \\
\midrule
PostgreSQL & $4.73\mathrm{E}{-01}$ & $7.42\mathrm{E}{-02}$ & $2.80\mathrm{E}{+00}$ & $7.18\mathrm{E}{-02}$ & $1.90\mathrm{E}{+01}$ \\
DuckDB & $1.03\mathrm{E}{-01}$ & $1.01\mathrm{E}{-01}$ & \scalabilitysecond{$1.65\mathrm{E}{-01}$} & $7.82\mathrm{E}{-02}$ & $1.37\mathrm{E}{-01}$ \\
ArangoDB & $4.00\mathrm{E}{-01}$ & $3.16\mathrm{E}{+00}$ & $3.44\mathrm{E}{-01}$ & $3.94\mathrm{E}{-01}$ & $3.39\mathrm{E}{-01}$ \\
AgensGraph & $1.11\mathrm{E}{-01}$ & \scalabilitysecond{$5.76\mathrm{E}{-02}$} & $1.15\mathrm{E}{+00}$ & $4.54\mathrm{E}{-02}$ & $1.74\mathrm{E}{+01}$ \\
OrientDB & \scalabilitybest{$2.05\mathrm{E}{-02}$} & $1.66\mathrm{E}{-01}$ & $2.03\mathrm{E}{-01}$ & \scalabilitybest{$1.24\mathrm{E}{-02}$} & \scalabilitysecond{$2.86\mathrm{E}{-02}$} \\
GRainDB & -- & -- & -- & -- & -- \\
K\`{u}zu & \scalabilitysecond{$3.66\mathrm{E}{-02}$} & \scalabilitybest{$2.33\mathrm{E}{-02}$} & \scalabilitybest{$1.62\mathrm{E}{-01}$} & \scalabilitysecond{$3.32\mathrm{E}{-02}$} & \scalabilitybest{$1.44\mathrm{E}{-02}$} \\

\bottomrule
\end{tabular}
\vspace{-2.25em}
\end{table}

\begin{table*}[t]
    \centering
    \small
    \setlength{\tabcolsep}{3.2pt}
    \renewcommand{\arraystretch}{0.95}
    \caption{Update latency at $SF=10^2$ (seconds). The task templates are grouped by update type: U1--U2 are vertex-only updates, U3--U4 update edges, and U5--U6 update vertices with their incident edges. Each template has deletion and insertion variants. Dark and light green denote the fastest and second-fastest results, respectively. ``--'' denotes unavailable measurements.}
    \vspace{-1.25em}
    \label{tab:exp update}
    \definecolor{updatebestgreen}{RGB}{180,205,180}
    \definecolor{updatesecondgreen}{RGB}{220,232,220}
    \newcommand{\updatebest}[1]{\cellcolor{updatebestgreen}#1}
    \newcommand{\updatesecond}[1]{\cellcolor{updatesecondgreen}#1}
    \resizebox{\textwidth}{!}{%
    \begin{tabular}{@{}c|cc|cc|cc|cc|cc|cc@{}}
    \toprule
    \multirow{2}{*}{\textbf{Systems}} & \multicolumn{6}{c|}{\textbf{Deletion}} & \multicolumn{6}{c}{\textbf{Insertion}} \\
    \cmidrule(lr){2-7}\cmidrule(lr){8-13}
     & \textbf{U1} & \textbf{U2} & \textbf{U3} & \textbf{U4} & \textbf{U5} & \textbf{U6} & \textbf{U1} & \textbf{U2} & \textbf{U3} & \textbf{U4} & \textbf{U5} & \textbf{U6} \\
    \midrule
PostgreSQL & -- & -- & $2.04\mathrm{E}{+02}$ & $5.05\mathrm{E}{+01}$ & $8.08\mathrm{E}{+02}$ & $8.16\mathrm{E}{+01}$ & -- & -- & $2.96\mathrm{E}{+01}$ & \updatesecond{$6.38\mathrm{E}{+00}$} & $1.53\mathrm{E}{+02}$ & \updatesecond{$8.37\mathrm{E}{+00}$} \\
    DuckDB & \updatebest{$1.68\mathrm{E}{+00}$} & \updatebest{$5.72\mathrm{E}{-01}$} & \updatesecond{$4.95\mathrm{E}{+00}$} & \updatesecond{$1.16\mathrm{E}{+00}$} & \updatebest{$1.64\mathrm{E}{+01}$} & \updatebest{$4.88\mathrm{E}{+00}$} & \updatesecond{$7.41\mathrm{E}{+00}$} & \updatebest{$3.88\mathrm{E}{+00}$} & \updatebest{$4.73\mathrm{E}{+00}$} & \updatebest{$3.14\mathrm{E}{+00}$} & \updatebest{$2.01\mathrm{E}{+01}$} & \updatebest{$5.49\mathrm{E}{+00}$} \\
    ArangoDB & $1.31\mathrm{E}{+02}$ & $2.59\mathrm{E}{+02}$ & $5.37\mathrm{E}{+02}$ & $4.38\mathrm{E}{+02}$ & $1.77\mathrm{E}{+03}$ & $8.73\mathrm{E}{+02}$ & $1.25\mathrm{E}{+02}$ & $2.39\mathrm{E}{+02}$ & $2.66\mathrm{E}{+02}$ & $3.42\mathrm{E}{+02}$ & $1.15\mathrm{E}{+03}$ & $5.92\mathrm{E}{+02}$ \\
AgensGraph & -- & -- & $2.13\mathrm{E}{+03}$ & $1.54\mathrm{E}{+03}$ & $1.80\mathrm{E}{+04}$ & $5.86\mathrm{E}{+03}$ & -- & -- & $1.19\mathrm{E}{+03}$ & $8.22\mathrm{E}{+02}$ & $1.71\mathrm{E}{+04}$ & $3.67\mathrm{E}{+03}$ \\
    OrientDB & $1.80\mathrm{E}{+04}$ & $4.02\mathrm{E}{+03}$ & $5.19\mathrm{E}{+03}$ & $1.31\mathrm{E}{+03}$ & $1.80\mathrm{E}{+04}$ & $1.60\mathrm{E}{+03}$ & $6.26\mathrm{E}{+03}$ & $1.29\mathrm{E}{+03}$ & $2.75\mathrm{E}{+03}$ & $9.56\mathrm{E}{+02}$ & $6.45\mathrm{E}{+03}$ & $9.31\mathrm{E}{+02}$ \\
    GRainDB & \updatesecond{$3.89\mathrm{E}{+01}$} & \updatesecond{$4.00\mathrm{E}{+01}$} & $7.53\mathrm{E}{+01}$ & $6.31\mathrm{E}{+01}$ & $3.53\mathrm{E}{+02}$ & $1.90\mathrm{E}{+02}$ & \updatebest{$5.88\mathrm{E}{+00}$} & \updatesecond{$6.09\mathrm{E}{+00}$} & \updatesecond{$1.21\mathrm{E}{+01}$} & $9.13\mathrm{E}{+00}$ & \updatesecond{$7.10\mathrm{E}{+01}$} & $4.61\mathrm{E}{+01}$ \\
    K\`{u}zu & $2.74\mathrm{E}{+03}$ & $8.51\mathrm{E}{+01}$ & \updatebest{$7.83\mathrm{E}{-01}$} & \updatebest{$2.31\mathrm{E}{-01}$} & \updatesecond{$3.19\mathrm{E}{+02}$} & \updatesecond{$5.14\mathrm{E}{+01}$} & $1.47\mathrm{E}{+01}$ & $6.29\mathrm{E}{+00}$ & $2.34\mathrm{E}{+01}$ & $1.97\mathrm{E}{+01}$ & $1.02\mathrm{E}{+02}$ & $2.90\mathrm{E}{+01}$ \\

    \bottomrule
    \end{tabular}}
    \end{table*}

\begin{figure*}[t]
    \centering
    \vspace{-1.25em}
    \includegraphics[width=\textwidth]{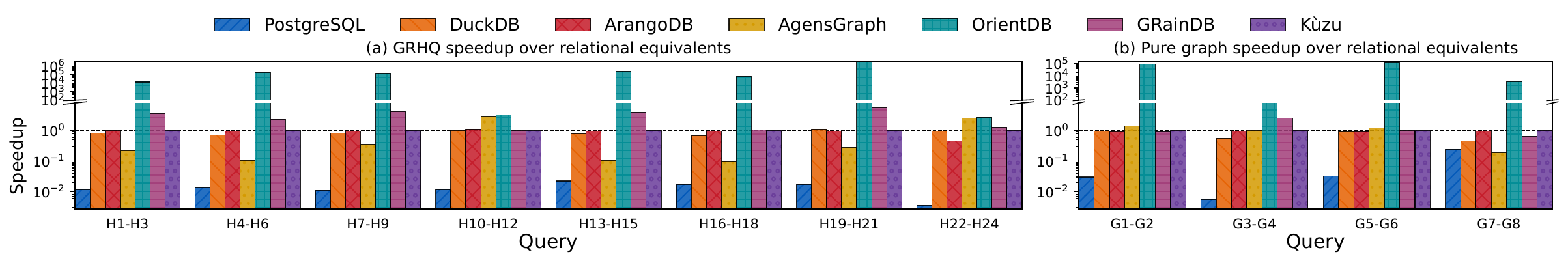}
    \vspace{-2.75em}
    \caption{Speedup over relational equivalents for H- and G-series queries.}
    \vspace{-1.5em}
    \label{fig:exp relational}
\end{figure*}

\subsection{Graph Operator Scalability (RQ2)}
\label{sec:exp scalability}
Table~\ref{tab:scalability-s-sf10-1} reports the geometric-mean latency across the three predicate-selectivity variants of each S-series query at $SF=10^1$. Observations:
(1) GRainDB is the only system unable to execute any S-series query. GRainDB supports variable-length path patterns through recursive common table expressions but does not implement the shortest-path semantics required by the S-series. This shows that efficient pattern matching does not necessarily imply extensibility to additional graph operators.
(2) No implementation dominates all S1--S5 queries at $SF=10^1$: K\`{u}zu achieves the lowest latency on S2, S3, and S5, while OrientDB performs best on S1 and S4. Both systems provide native path operators, while DuckPGQ uses an on-demand CSR representation with vectorized path search, and AGE and AgensGraph rely on PostgreSQL-backed path expansion. Although these designs all support shortest-path processing at this scale, their different execution mechanisms may exhibit different scaling behavior as the graph grows, which we examine in Section~\ref{sec:exp data scale}.

\subsection{Pure Graph Query Performance (RQ3)}\label{sec:exp graph}
We evaluate all systems on the 8 pure-graph G-series queries at $SF=10^1$ and report their latencies in Figure~\ref{fig:exp graph}. Observations:
(1) DuckDB achieves the lowest latency across most G-series queries. This is because DuckPGQ lowers SQL/PGQ graph patterns into DuckDB's relational execution framework, allowing graph matching to reuse its vectorized parallel execution engine without a separate graph execution layer.
(2) PostgreSQL is competitive on selected queries such as G1, G5, and G8, but becomes substantially slower on G2--G4 and G6--G7. For multi-subgraph queries, AGE processes multiple graph matching results through the \texttt{cypher()} interface before aggregation, increasing overhead as intermediate results grow.
(3) OrientDB performs well on selected anchored queries such as G7 and G8, but becomes less competitive on queries such as G3 and G4. Its traversal execution benefits from selective starting points and direct navigation, but provides less advantage when graph patterns produce larger intermediate results or multiple results must be combined.
(4) ArangoDB and GRainDB show relatively stable but moderate performance across G-series structures. This indicates that graph-native storage or predefined access structures alone do not determine latency, which also depends on starting predicates and intermediate result sizes.
(5) Hop depth and matched-subgraph count do not cause monotonic latency growth. For example, G6 has a deeper pattern than G5 but does not incur proportionally higher latency, while G2 is much slower than G1 despite sharing the same structural category. Thus, intermediate result cardinality and execution strategy can matter more than hop count alone.

\subsection{Graph Update Performance (RQ4)}\label{sec:exp update}

We evaluate the paired U-series deletion and insertion tasks at $SF=10^2$ and report their query-level latencies in Table~\ref{tab:exp update}. We make the following observations:
(1) DuckDB is fastest on most insertion and deletion tasks. This is because it defines a property graph directly over its vertex and edge tables, so the measured graph updates are ordinary relational \texttt{INSERT} and \texttt{DELETE} operations on the underlying tables and require no synchronization with a separate graph copy.
(2) K\`{u}zu and GRainDB achieve the best or second-best latency on many U-series tasks. In particular, GRainDB consistently outperforms ArangoDB, AgensGraph, and OrientDB across all measured deletion and insertion tasks. K\`{u}zu stores graph topology in native CSR-based adjacency structures, allowing topology updates to operate directly on its graph storage. GRainDB reuses DuckDB's table-update path but additionally maintains materialized RID references and optional RID indexes, making its deletion latencies $4.1$--$6.9\times$ higher than the corresponding insertion latencies.
(3) ArangoDB, AgensGraph, and OrientDB generally exhibit substantially higher update latencies, with AgensGraph and OrientDB frequently forming the slowest tier. ArangoDB maintains edge collections and the automatic indexes on their \texttt{\_from} and \texttt{\_to} fields, while OrientDB removes edge references from both endpoint vertices. AgensGraph stores vertices and edges in PostgreSQL label tables and enforces consistency between them. These additional index and topology maintenance operations contribute to their higher bulk update costs.
(4) PostgreSQL with AGE and AgensGraph do not support the standalone vertex deletions in U1 and U2 because both systems reject deleting vertices that still have incident edges.

\begin{figure*}[t]
    \centering
    \includegraphics[width=\textwidth]{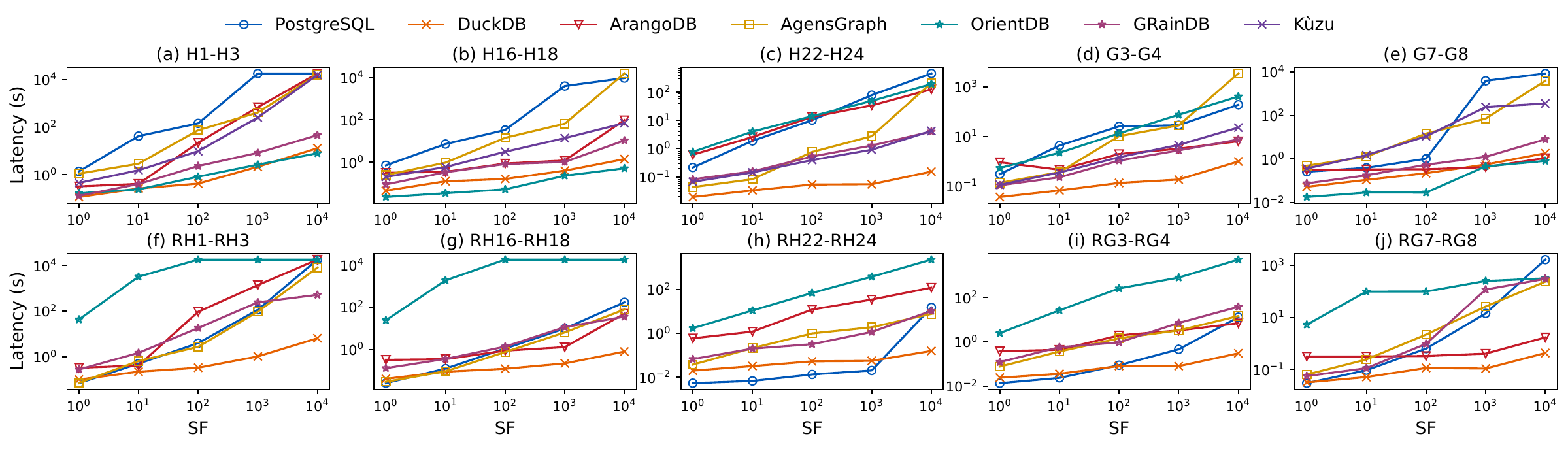}
    \vspace{-2.75em}
    \caption{Impact of data scale on representative H- and G-series query groups and their relational equivalents.}
    \vspace{-1.5em}
    \label{fig:exp sf}
\end{figure*}

\begin{figure}[t]
    \centering
    \includegraphics[width=0.48\textwidth]{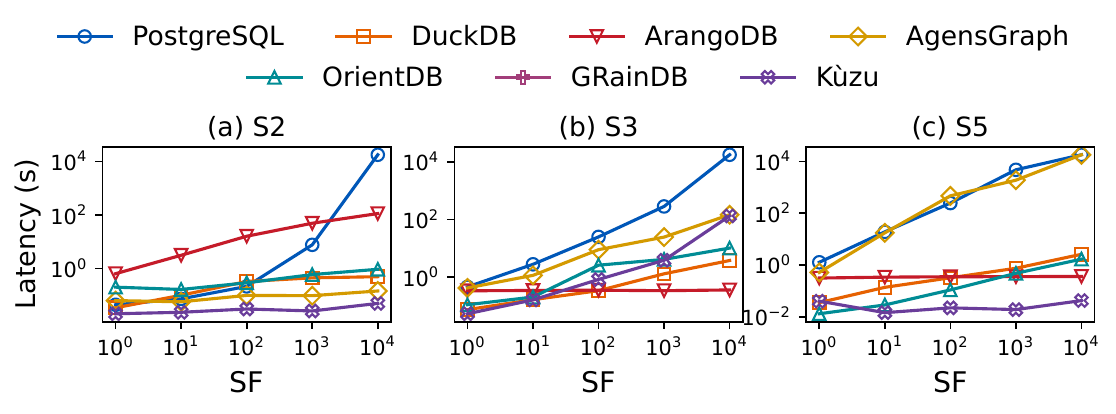}
    \vspace{-3.25em}
    \caption{Impact of data scale on representative S-series queries (seconds).}
    \vspace{-1.5em}
    \label{fig:exp s sf}
\end{figure}

\subsection{Speedup over Relational Equivalents (RQ5)}\label{sec:exp relational}

To quantify graph-oriented execution benefits, we compare each H- and G-series query with its equivalent relational formulation in the R-series at $SF=10^1$. For each group, Figure~\ref{fig:exp relational} reports the geometric mean of the graph-relational speedup over its queries and three selectivity variants. We make the following observations:
(1) PostgreSQL with AGE remains below the break-even line for every H- and G-series group. AGE translates Cypher into PostgreSQL plans over vertex and edge label tables and represents graph values with \texttt{agtype}. Fixed-depth patterns incur Cypher translation and graph-type handling while still executing on relational storage, whereas the relational equivalents directly join typed identifiers. The G-series results show that the disadvantage stems from AGE's graph representation and \texttt{MATCH} execution, not only from composing \texttt{cypher()} outputs with SQL.
(2) DuckDB and ArangoDB remain close to, but mostly below, the break-even line. DuckPGQ lowers SQL/PGQ patterns into plans over the same DuckDB base tables as the relational equivalents, so the graph formulation gains no separate graph access path while incurring additional lowering and extension-processing overhead. In ArangoDB, native traversal and join-based AQL both exploit the automatically maintained \texttt{\_from} and \texttt{\_to} indexes for endpoint lookup, so the shallow, fixed-depth patterns incur similar access costs and exhibit comparable latency.
(3) OrientDB exceeds the break-even line across nearly all H- and G-series groups, while GRainDB does so in many groups and AgensGraph only in selected cases. OrientDB's \texttt{MATCH} executor can select indexed starting vertices before navigating graph links, while GRainDB replaces joins on identifier values with predefined RID joins and optional RID indexes. These graph-aware access paths avoid the repeated identifier-based joins required by the relational translations. In the inspected AgensGraph plans, graph query results are materialized before outer joins, and one intermediate sort spills to disk, which limits its speedup for some query structures.
(4) K\`{u}zu exposes only a Cypher-based graph query interface and has no semantically equivalent relational formulation. We therefore set its speedup to 1 as a reference value.

\subsection{Impact of Data Scale}
\label{sec:exp data scale}

\myparagraph{Impact of Data Scale on H-, G-, and R-series}
We evaluate scalability by varying the scale factor from $SF=10^0$ to $SF=10^4$ and report latency trends of representative H-, G-, and R-series workloads in Figure~\ref{fig:exp sf}. At each scale, the first row reports geometric mean latency for each selected H- or G-series group, while the second reports that of corresponding relational equivalents. We observe that:
(1) DuckDB shows the most stable scaling across workloads. This is consistent with DuckPGQ lowering graph patterns into DuckDB's relational execution framework, allowing graph matching and relational operators to scale through the same execution engine.
(2) PostgreSQL shows steep latency growth for H-series queries as scale increases, especially on H1--H3 and H16--H18. The increase mainly occurs in workloads with larger joins and graph-derived intermediate results, where AGE must materialize Cypher results before combining them with PostgreSQL relational execution.
(3) Graph and relational formulations can exhibit substantially different scaling behavior. OrientDB's H- and G-series queries remain competitive on several workloads, whereas their relational equivalents grow rapidly with scale. In contrast, GRainDB maintains relatively stable growth on graph queries by exploiting predefined relationships and RID-based access paths.
(4) Increasing scale does not uniformly amplify the effect of hop depth or query structure. For example, some multi-hop queries scale better than shorter queries when intermediate results remain limited. Therefore, scalability is mainly determined by how systems handle intermediate graph results and relational composition rather than hop count alone.

\myparagraph{Impact of Data Scale on S-series}
Figure~\ref{fig:exp s sf} shows how data scale affects geometric mean latency across three predicate selectivity variants of representative S-series queries. Observations:
(1) PostgreSQL exhibits the sharpest latency growth across S2, S3, and S5, while AgensGraph grows sharply on S3 and S5 but remains stable on S2. Their PostgreSQL-backed path execution can expand increasing traversal states and intermediate results as the graph grows.
(2) ArangoDB scales well on S3 and S5 but deteriorates markedly on S2. This is because S2 searches from a fixed source toward predicate-selected destinations, whereas ArangoDB's shortest-path execution uses explicit source--destination traversal, causing the growing destination set to amplify repeated path-search work.
(3) DuckDB remains comparatively stable across all three queries. DuckPGQ constructs an on-demand CSR representation and executes shortest-path search with vectorized graph algorithms, limiting repeated relational or traversal overhead as the graph expands.

\subsection{Impact of Predicate Selectivity}
\label{sec:exp selectivity}
\begin{table}[t]
    \centering
    \scriptsize
    \setlength{\tabcolsep}{1.8pt}
    \renewcommand{\arraystretch}{0.95}
    \caption{Relative predicate-selectivity response for representative H-, G-, and S-series queries at $SF=10^1$. 
    }
    \vspace{-1.5em}
    \label{tab:exp selectivity}
    \definecolor{selectivitybestgreen}{RGB}{180,205,180}
    \definecolor{selectivitysecondgreen}{RGB}{220,232,220}
    \newcommand{\selectivitybest}[1]{\cellcolor{selectivitybestgreen}#1}
    \newcommand{\selectivitysecond}[1]{\cellcolor{selectivitysecondgreen}#1}
    \resizebox{\columnwidth}{!}{%
    \begin{tabular}{@{}c@{\hspace{2pt}}|@{\hspace{2pt}}*{2}{c}@{\hspace{2pt}}|@{\hspace{2pt}}*{2}{c}@{\hspace{2pt}}|@{\hspace{2pt}}*{2}{c}@{\hspace{2pt}}|@{\hspace{2pt}}*{2}{c}@{\hspace{2pt}}|@{\hspace{2pt}}*{2}{c}@{}}
    \toprule
    \textbf{Systems} & \textbf{H3b} & \textbf{H3c} & \textbf{H10b} & \textbf{H10c} & \textbf{G1b} & \textbf{G1c} & \textbf{G4b} & \textbf{G4c} & \textbf{S3b} & \textbf{S3c} \\
    \midrule
    PostgreSQL & 0.88 & \selectivitybest{0.07} & \selectivitybest{0.78} & \selectivitysecond{0.64} & 0.95 & \selectivitysecond{0.81} & 0.96 & 0.50 & 0.99 & 1.01 \\
    DuckDB & 0.99 & 0.99 & 0.97 & \selectivitybest{0.63} & 1.29 & 1.18 & 0.91 & 0.70 & \selectivitysecond{0.61} & 0.60 \\
    ArangoDB & 0.99 & 0.98 & 1.00 & 0.99 & 1.00 & 1.00 & 0.99 & 0.99 & 0.96 & 0.87 \\
    AgensGraph & \selectivitysecond{0.86} & 0.85 & \selectivitysecond{0.88} & 0.70 & \selectivitysecond{0.93} & 0.92 & \selectivitysecond{0.88} & 0.61 & \selectivitybest{0.54} & \selectivitybest{0.02} \\
    OrientDB & \selectivitybest{0.25} & \selectivitysecond{0.50} & 0.99 & 1.07 & \selectivitybest{0.69} & \selectivitybest{0.68} & \selectivitybest{0.09} & \selectivitybest{0.02} & 0.96 & \selectivitybest{0.02} \\
    GRainDB & 1.00 & 0.96 & 1.00 & 0.83 & 0.97 & 3.54 & 1.00 & \selectivitysecond{0.18} & -- & -- \\
    K\`{u}zu & 1.10 & 0.87 & 0.96 & 0.95 & 1.06 & 0.95 & 1.03 & 1.03 & 0.83 & 0.81 \\
    \bottomrule
    \end{tabular}}
    \vspace{-1.5em}
    \end{table}

We evaluate the effect of predicate selectivity using representative H-, G-, and S-series queries with three selectivity variants. Specifically, for each query $Q$, we use $\mathcal{L}(Q_b)/\mathcal{L}(Q_a)$ and $\mathcal{L}(Q_c)/\mathcal{L}(Q_a)$ to quantify each system's sensitivity to selectivity changes, where $Q_a$, $Q_b$, and $Q_c$ denote the three variants of $Q$ with progressively lower predicate selectivities. Table~\ref{tab:exp selectivity} reports the resulting latency ratios. We make the following observations:
(1) OrientDB shows the strongest response to predicate selectivity on several graph queries, with G4b and G4c reducing latency to $0.09$ and $0.02$ of their baseline values. PostgreSQL and AgensGraph also benefit substantially on selected queries, such as H3c and S3c, when selective predicates can be applied before graph expansion or pushed into scans and joins, thereby reducing starting vertices or intermediate results. In contrast, ArangoDB, most K\`{u}zu cases, and several DuckDB cases remain close to one. In these queries, stricter predicates do not substantially reduce the starting set or prune graph expansion early, so similar scans and fixed-depth patterns still dominate execution.
(2) G1b and G1c add an endpoint-property predicate absent from G1a, introducing additional property access and filtering. GRainDB reaches $3.54$ on G1c, suggesting that the additional predicate changes the placement of filtering or join evaluation in its predefined relational access path; the exact cause requires execution plans.
(3) S3 shows the strongest response in AgensGraph and OrientDB under the most selective variant, moderate sensitivity in DuckDB and K\`{u}zu, and little change in PostgreSQL and ArangoDB. S3 places the predicate on the terminal vertex of a three-hop citation pattern before joining the graph result with metadata relations. Systems that push it into traversal or graph result production can prune intermediates early, whereas broad fixed-depth scans or similar materialization gain less.

\subsection{Query Language Conciseness}
\label{sec:exp language}

\begin{table}[t]
    \centering
    \scriptsize
    \setlength{\tabcolsep}{2.4pt}
    \renewcommand{\arraystretch}{0.95}
    \caption{Language conciseness evaluation.}
    
    \vspace{-1.5em}
    \label{tab:exp language}
    \definecolor{languagebestgreen}{RGB}{180,205,180}
    \definecolor{languagesecondgreen}{RGB}{220,232,220}
    \newcommand{\languagebest}[1]{\cellcolor{languagebestgreen}#1}
    \newcommand{\languagesecond}[1]{\cellcolor{languagesecondgreen}#1}
    \resizebox{\columnwidth}{!}{%
    \begin{tabular}{@{}c|ccc|ccc@{}}
    \toprule
    \multirow{2}{*}{\textbf{Systems}} & \multicolumn{3}{c|}{\textbf{H}} & \multicolumn{3}{c}{\textbf{G}} \\
    \cmidrule(lr){2-4}\cmidrule(lr){5-7}
     & \textbf{Volume} & \textbf{Difficulty} & \textbf{Effort} & \textbf{Volume} & \textbf{Difficulty} & \textbf{Effort} \\
    \midrule
    PostgreSQL & 1.11E+03 & 2.94E+01 & 3.43E+04 & 5.95E+02 & 2.52E+01 & 1.69E+04 \\
    DuckDB & 9.95E+02 & 2.81E+01 & 2.89E+04 & 5.42E+02 & 2.38E+01 & 1.41E+04 \\
    ArangoDB & 1.33E+03 & 3.41E+01 & 4.60E+04 & \languagebest{3.46E+02} & \languagebest{1.89E+01} & \languagebest{7.67E+03} \\
    AgensGraph & 1.22E+03 & 2.88E+01 & 3.68E+04 & 6.54E+02 & 2.51E+01 & 1.86E+04 \\
    OrientDB & 2.09E+03 & 4.25E+01 & 9.68E+04 & 7.27E+02 & 2.58E+01 & 2.19E+04 \\
    GRainDB & \languagebest{8.43E+02} & \languagebest{2.66E+01} & \languagebest{2.32E+04} & \languagesecond{4.03E+02} & \languagesecond{2.02E+01} & \languagesecond{9.67E+03} \\
    K\`{u}zu & \languagesecond{9.81E+02} & \languagebest{2.66E+01} & \languagesecond{2.64E+04} & 4.63E+02 & 2.03E+01 & 9.86E+03 \\
    \bottomrule
    \end{tabular}}
    \vspace{-1.5em}
    \end{table}

We compare the complexity of H and G query formulations and present their geometric mean in Table~\ref{tab:exp language}. Lower values indicate more concise query expressions. Observations:
(1) GRainDB is the most concise for H-series queries, whereas OrientDB has the highest complexity. GRainDB compactly expresses graph patterns and composes bindings with SQL joins, while OrientDB nests \texttt{MATCH}, \texttt{LET}, and \texttt{IN} clauses for multiple graph outputs and joins, increasing both operator and operand counts.
(2) ArangoDB is the most concise for G-series queries. Pure graph patterns map directly to compact AQL traversal constructs, whereas H-series formulations require additional \texttt{LET} subqueries and explicit join logic to compose graph results, increasing complexity.

\subsection{Resource Utilization}
\label{sec:exp resource}

\begin{table}[t]
\centering
\scriptsize
\setlength{\tabcolsep}{2.2pt}
\renewcommand{\arraystretch}{1.06}
\caption{Resource utilization (CPU: mean; MEM: peak).}
\vspace{-1.5em}
\label{tab:exp resource}
\resizebox{\columnwidth}{!}{%
\begin{tabular}{@{}cccccc@{}}
\toprule
\textbf{System} & \textbf{H (CPU, MEM)} & \textbf{G (CPU, MEM)} & \textbf{S (CPU, MEM)} & \textbf{U (CPU, MEM)} & \textbf{R (CPU, MEM)}\\
\midrule
PostgreSQL & $(2.45,\,47.9)$ & $(1.73,\,37.2)$ & $(0.98,\,73.7)$ & $(0.95,\,48.7)$ & $(0.92,\,6.39)$ \\
DuckDB & $(58.6,\,3.49)$ & $(55.9,\,2.96)$ & $(4.60,\,0.69)$ & $(8.16,\,3.05)$ & $(57.0,\,2.40)$ \\
ArangoDB & $(0.12,\,41.4)$ & $(0.13,\,41.4)$ & $(0.18,\,41.7)$ & $(1.04,\,25.4)$ & $(0.13,\,41.0)$ \\
AgensGraph & $(3.35,\,25.1)$ & $(3.35,\,25.2)$ & $(0.65,\,0.11)$ & $(0.30,\,30.0)$ & $(3.07,\,0.92)$ \\
OrientDB & $(1.84,\,6.70)$ & $(1.40,\,11.9)$ & $(1.86,\,12.3)$ & $(1.72,\,64.9)$ & $(3.88,\,25.2)$ \\
GRainDB & $(0.85,\,31.5)$ & $(0.68,\,33.3)$ & -- & $(0.62,\,71.6)$ & $(0.90,\,29.8)$ \\
K\`{u}zu & $(6.01,\,10.7)$ & $(5.03,\,11.2)$ & $(1.77,\,1.29)$ & $(2.74,\,30.6)$ & -- \\
\bottomrule
\end{tabular}}
\vspace{-2em}
\end{table}

We measure the mean CPU utilization and peak memory usage across the five GRBench workload series at $SF=10^2$ and report the geometric mean in Table~\ref{tab:exp resource}. We make the following observations:
(1) DuckDB consumes higher CPU resources than most systems, especially on H and G workloads. This higher utilization corresponds to its parallel execution model and is consistent with its low query latency.
(2) PostgreSQL uses substantially more memory on H, G, and S than on R, while ArangoDB remains near 41 GB across read workloads. AGE carries Cypher results as \texttt{agtype} records, increasing memory pressure from graph intermediates, whereas ArangoDB's RocksDB engine retains a hot set of data and indexes in memory, so cache allocation dominates its stable footprint.
(3) Update workloads produce the highest memory peaks for OrientDB and GRainDB. OrientDB updates edge records together with references at both endpoint vertices, while GRainDB maintains base tuples, materialized RIDs, and RID indexes, causing batch updates to retain more topology state in memory. 

\section{Open Problems}
Based on our experimental results, we identify the following five open problems for graph-relational data management:

\begin{itemize}[left=0pt]

\item \textbf{Joint Optimization of Graph and Relational Operators.}
H-series performance varies substantially with join count, matched subgraphs, and intermediate-result size, exposing the limitations of optimizing graph and relational operators in isolation. A key challenge is to develop unified cardinality and cost models with plan-search strategies that jointly optimize graph pattern evaluation and graph-relational join orders.

\item \textbf{Extensible Graph Operator Integration.}
The S-series shows efficient pattern matching does not necessarily extend to shortest-path processing. Graph operators therefore need common algebraic representations, plan organizations, statistics, and physical interfaces, allowing optimizers to compose, transform, and cost new operators without per-operator redesign.

\item \textbf{Selectivity-aware Optimization.}
Predicate selectivity causes large, sometimes non-monotonic performance differences across systems. This motivates graph-aware cardinality estimation that captures how predicate effects propagate through starting vertices, graph expansion, joins, and intermediate results, enabling better predicate placement and traversal pruning.

\item \textbf{Adaptive Graph vs. Relational Execution.}
Our relational-equivalent comparison shows graph execution is not consistently faster than relational joins. Optimizers should treat graph operators and relational joins as alternative physical implementations and choose based on query and data characteristics.

\item \textbf{Update-oriented Graph-relational Storage and Consistency.}
The U-series shows that update cost and support depend on topology representation, from table-backed graphs to RID- and adjacency-based structures. An open problem is to maintain consistency across base records, topology structures, and auxiliary indexes while preserving update and query efficiency.

\end{itemize}
\section{Conclusions}
In this work, we present \bench, a comprehensive benchmark for graph-relational data management. \bench combines a scalable real-world graph-relational dataset, purpose-built query series, and semantically equivalent native formulations to systematically evaluate representative database systems. Our evaluation reveals substantial differences across storage, execution, and optimization designs, with no single architecture consistently dominating all workloads. These results highlight the trade-offs of existing approaches and provide guidance for future graph-relational system design and optimization.

\bibliographystyle{ACM-Reference-Format}
\bibliography{sample-base}

\end{document}